\documentclass[longbibliography,
twocolumn,prl]{revtex4-2}
\usepackage[T1]{fontenc}
\usepackage[utf8]{inputenc}
\usepackage[a4paper]{geometry}
\usepackage{color}
\usepackage{amsmath}
\usepackage{amsfonts,amssymb,txfonts}
\usepackage{epsfig,color}
\usepackage{graphicx}
\usepackage{hyperref}
\usepackage{float}
\usepackage{comment}
\usepackage{placeins}
\hypersetup{colorlinks=true,linkcolor=blue,anchorcolor=blue,citecolor=blue,filecolor=blue,urlcolor=blue,bookmarksnumbered=true,pdfview=FitB}

\newcommand{\ve}{\varepsilon}

\newcommand{\bk}{{\bf k}}
\newcommand{\bp}{{\bf p}}

\newcommand{\bq}{{\bf q}}

\newcommand{\nn}{\nonumber}
\newcommand{\beq}{\begin{equation}}
\newcommand{\eeq}{\end{equation}}
\newcommand{\bea}{\begin{eqnarray}}
\newcommand{\eea}{\end{eqnarray}}
\newcommand{\bse}{\begin{subequations}}
\newcommand{\ese}{\end{subequations}}
\newcommand{\bwt}{\begin{widetext}}
\newcommand{\ewt}{\end{widetext}}
\newcommand\re{{\mathrm{Re}}}
\newcommand\im{{\mathrm{Im}}}
\newcommand{\bv}{{\bf v}}
\newcommand{\bu}{{\bf u}}
\newcommand{\br}{{\bf r}}

\newcommand{\bsu}{\begin{subequations}}
\newcommand{\esu}{\end{subequations}}

\newcommand{\bQ}{{\bf Q}}
\newcommand{\e}{\epsilon}
\newcommand{\kf}{k_\mathrm{F}}
\newcommand{\vf}{\varv_\mathrm{F}}

\newcommand{\ef}{\varepsilon_{\text{F}}}
\newcommand{\nf}{\nu_{\text{F}}}

\newcommand{\bi}{\begin{itemize}}
\newcommand{\ei}{\end{itemize}}

\newcommand{\ts}{\tau_{\mathrm{s}}}
\newcommand{\sgs}{\sigma_{\mathrm{s}}}
\newcommand{\rs}{\rho_{\mathrm{s}}}
\newcommand{\Ds}{D_{\mathrm{s}}}

\newcommand{\ls}{\ell_{\mathrm{s}}}
\newcommand{\sigs}{\sigma_{\rm s}}

\makeatletter

\@ifundefined{date}{}{\date{\today}}
\makeatother

\usepackage{babel}
\usepackage{graphics}
\begin{document}
\title{Resistive Anomaly near a 
Ferromagnetic
Phase Transition:\\
A Classical Memory Effect}
\author{
Dmitrii L. Maslov$^{1}$}
\author{Vladimir I. Yudson$^{2,3}$}
\author{Cristian D. Batista$^{4,5}$}
\affiliation{$^{1}$Department of Physics, University of Florida, P.O. Box Gainesville,
Florida 32611}
\affiliation{$^{2}$Laboratory for Condensed Matter Physics, HSE University, 20 Myasnitskaya St., Moscow, 101000 Russia}
\affiliation{$^{3}$ Russian Quantum Center, Skolkovo, Moscow 143025, Russia}
\affiliation{$^{4}$ Department of Physics and Astronomy, University of Tennessee, Knoxville, TN 37996, USA}
\affiliation{$^{5}$ Neutron Scattering Division, Oak Ridge National Laboratory, Oak Ridge, TN 37831, USA}
\begin{abstract}

 We investigate resistive anomalies in metals near a ferromagnetic phase transition,
 emphasizing the role of long-range critical fluctuations. Our analysis shows that electron diffusion near the critical temperature $T_c$ enhances the singular behavior of resistivity via a classical memory effect, exceeding the prediction of Fisher and Langer~\cite{Fisher:1968}. Close to $T_c$, the resistivity develops a cusp or anticusp 
 controlled
 by the critical exponent of the order parameter. We also express 
 a
 concomitant non-Drude behavior of the optical conductivity in terms of critical exponents. These results provide deeper insight into the origin of resistive anomalies and their connection to criticality in metallic systems.

\end{abstract}
\maketitle
\paragraph{Introduction:}

Second-order phase transitions 
are accompanied by diverging
order-parameter fluctuations near the critical temperature $T_c$. A classic manifestation is critical opalescence—
strong scattering of light by a fluid near its critical point. In metals approaching a magnetic or structural transition, electron waves are similarly scattered by critical fluctuations. The metallic analog of critical opalescence is a knee- or cusp-like anomaly in the temperature dependence of resistivity near $T_c$.


Since the pioneering work of Gerlach~\cite{Gerlach1932}, resistive anomalies have been widely observed in elemental metallic ferromagnets (FMs), such as Ni, Fe, Co, and Gd~\cite{Campbell:1982}, in rare-earth intermetallics~\cite{Mydosh:1970,Blanco1994}, and in metallic antiferromagnets (AFMs), including elemental (Tb, Ho)~\cite{Nellis:1968}, intermetallic~\cite{Fote:1970}, and iron pnictides~\cite{Hristov:2019}. Similar anomalies also occur near structural transitions, e.g., near order-disorder transitions in binary alloys~\cite{simons:1971} and the cubic-to-tetragonal transition in doped SrTiO$_3$~\cite{Sugawara1:988}. 


Resistive anomalies offer two main advantages: they enable rapid identification of magnetic transitions without direct magnetic measurements, and their shapes reveals information about critical exponents. This is particularly valuable given the higher precision of resistivity measurements compared to specific heat. However, extracting such information requires a suitable theoretical framework~\cite{Kallback81}, which we revisit and refine in this Letter.

Theoretical studies of resistive anomalies began with de Gennes and Friedel (dGF)~\cite{deGennes:1958}, who attributed the 
resistive anomaly
near a FM transition to spin-flip scattering of free electrons by localized magnetic moments. Modeling this as quenched long-range disorder (LRD), they computed the transport scattering time $\tau_{\rm tr}$, using the Fermi's Golden Rule (FGR):
\bea
1/\tau_{\rm tr} \propto \int_{q\leq 2k_F} d^d q \, \frac{q^2}{2\kf^2} \, \Delta t(q) \, W(\bq), \label{FGR}
\eea
where $\kf$ is the Fermi momentum, $W(\bq) = \langle \delta {\bf S}_{\bq} \cdot \delta {\bf S}_{-\bq} \rangle$ is the connected spin correlation function, $\Delta t(q) = 1/\vf q$ 
arises from the energy-conserving
delta-function
averaged over 
directions of $\bq$,
and the ``transport factor''--$q^2/2\kf^2$--suppresses small-angle scattering. In the mean-field theory, $W(q) \propto \left((\bq - \bQ)^2 + \xi^{-2}\right)^{-1}$, 
where the correlation length $\xi \propto |\theta|^{-1/2}$  and $\theta=(T-T_c)/T_c$. 
For FMs, 
$\bQ = 0$ 
and thus Eq.~\eqref{FGR} is dominated by  $q \sim \kf$.
Subtracting off
this
non-universal part, dGF retained only the critical contribution from $q \sim \xi^{-1} \ll \kf$, leading to a resistivity cusp: $\delta\rho_{\rm dGF} \propto -|\theta| \ln (1/|\theta|)$ in $d=3$
and $\delta\rho_{\rm dGF} \propto -|\theta|^{1/2}$ in $d=2$.

Ten years later, Fisher and Langer (FL)~\cite{Fisher:1968} challenged two key aspects of dGF's analysis. First, they correctly pointed out that impurity and phonon scattering—
which plays the role pf short-range disorder (SRD)—cannot be treated independently of magnetic scattering,
if the corresponding mean free path $\ls$ is shorter than 
$\xi$; a condition inevitably met near $T_c$. While FL conjectured that this interplay would weaken the dGF singularity, they did not provide a quantitative treatment. Instead, they focused on the second issue: the short-range ($q \sim \kf$) contribution to 
Eq.~\eqref{FGR}, 
discarded by dGF, but which, 
in fact,
contributes to the critical $\theta$-dependence. 
Noting  that in typical metallic magnets $\kf \sim a^{-1}_{\rm M} \sim a^{-1}$, where $a_{\rm M}$ is the spacing between magnetic moments and $a$ the lattice constant, 
FL observed that the upper limit of the integral in Eq.~\eqref{FGR} probes the short-distance part of the spin correlation function. This contribution to $1/\tau_{\rm tr}$ scales with temperature 
as the magnetic internal energy, $U(T)$. Beyond the mean-field theory, the specific heat $C(T) = dU/dT \propto |\theta|^{-\alpha}$, with $0 < \alpha < 1$, leading FL to conclude that $d\delta\rho/dT \propto d\tau_{\rm tr}^{-1}/dT \propto C(T)$, or
\bea
\delta\rho_{\rm FL} \propto \text{sgn}\,\theta\, |\theta|^{1-\alpha}, \label{FL}
\eea
which is known as ``FL scaling''. Unlike the dGF prediction, FL scaling implies that $\rho(T)$ is a monotonic—generally increasing—function of $\theta$~\cite{Alexander:1976}, with a cusp in its derivative at $T_c$, consistent with observations in elemental FMs away from the critical point~\cite{Campbell:1982}.


 While FL’s argument applies to both FMs and AFMs, Suezaki and Mori (SM)~\cite{Suezaki:1969} pointed out an additional short-range contribution in metallic AFMs with $Q \sim \kf \sim a^{-1}$. In this case, the region $\bq \approx \bQ$ contributes critically to 
 the 
 integral~\eqref{FGR}, 
 while neither the transport factor nor $\Delta t(q)$ affect
 the scaling. 
 On changing variables to $\bq' = \bq - \bQ$, the integral is dominated by $q' \sim \xi^{-1}$, yielding
\bea
\delta\rho_{\rm SM} \propto \int dq'\, q'^{d-1} W(q') = \langle \delta
M^2 \rangle \propto |\theta|^{2\beta}, \label{SM}
\eea
where $\delta
M$ is the fluctuation of the magnetic order parameter
and $\beta$ is the critical exponent, defined by $\langle 
M\rangle \propto (-\theta)^\beta$ for $\theta < 0$.


Here, we show that in the diffusive regime ($\xi \gg \ls$), Eq.~\eqref{SM} applies not only to AFMs but also to FMs. In this case, the anomaly originates from long-range critical fluctuations—the same mechanism considered by dGF. Our result departs from dGF’s prediction due to the breakdown of Matthiessen’s rule, as noted by FL. Crucially, when properly accounted for, this breakdown enhances rather than suppresses the resistive anomaly. While Ref.~\cite{Timm:2005} attributed such enhancement to mesoscopic fluctuations, we demonstrate that even classical diffusive motion of phase-incoherent electrons in a background of long-range magnetic fluctuations produces a competing mechanism to FL scaling. The effect stems from a classical memory mechanism: repeated returns of a diffusive trajectory to the same location.

\paragraph{Qualitative arguments.}

As in previous work, we treat spin-flip scattering as arising from quenched long-range disorder (LRD), characterized by the correlation function $W(q) = \int d^d r\, e^{i\bq \cdot \br} \langle V(\br) V(0) \rangle$. This approximation is justified in the small-$q$ limit, where critical fluctuations exhibit critical slowing down: their relaxation time diverges as $q^{-z}$ near a continuous phase transition, with $z$ the dynamical exponent~\cite{Hohenberg:1977}. 
Assuming that LRD is weaker than 
SRD,
the key question 
is: what is the resulting resistivity of the metal?


Classical electrodynamics offers a partial answer to this question~\cite{Electrodynamics_LL}. In the diffusive regime, regions of size $\sim \xi$ act as local Ohmic resistors with spatially varying conductivity $\sigma(\br)$. Consequently, both the electric field and current density fluctuate, while obeying Ohm's law: ${\bf j}(\br) = \sigma(\br) {\bf E}(\br)$. 
The measured conductivity is defined via the 
relation between averaged quantities $\langle {\bf j} \rangle = \sigma_{\rm exp} \langle {\bf E} \rangle$, where $\langle {\bf E} \rangle$ is 
is the ratio of the voltage to sample length~\cite{Dykhne:1971}. For weak fluctuations, $\sigma_{\rm exp}$ splits into two contributions~\cite{Altshuler_JETP:1985}:
\bea
\sigma_{\rm exp} = \sigma_{\rm K} + \delta\sigma_{\rm fl}, \label{mes}
\eea
where $\sigma_{\rm K}$ is the Kubo conductivity in a uniform field, and $\delta\sigma_{\rm fl}$ is a correction due to spatial inhomogeneity. The result of Ref.~\cite{Electrodynamics_LL} for a weakly inhomogeneous medium
 translates into
 $\delta\sigma_{\rm fl} = -\langle \delta\sigma^2 \rangle / d \, \sigma_{\rm K}$~\cite{Dykhne:1971,Altshuler_JETP:1985,Timm:2005}. In our context, these conductivity fluctuations stem from order-parameter fluctuations, implying $\delta\sigma_{\rm fl} \propto \langle \delta M^2 \rangle$, thus justifying the scaling in Eq.~\eqref{SM}.


The Kubo conductivity 
itself consists of two parts: $\sigma_{\rm K} = \sigma_{\rm s} + \delta\sigma_{\ell}$, where $\sigma_{\rm s}$ is the Drude contribution from 
SRD
and $\delta\sigma_{\ell}$ is 
the correction due to 
LRD.
The form of $\delta\sigma_{\ell}$ depends on the relation between $\xi$ and $\ls$, with the key quantity being the ``interaction time'' $\Delta t(q)$ in Eq.~\eqref{FGR}, which 
is the duration of a scattering event with momentum transfer $q \sim \xi^{-1}$~\cite{altshuler:1985}. By the uncertainty principle, such momentum transfer occurs within a region of size $\sim 1/q$. 

In the ballistic regime ($\xi \ll \ls$), an electron traverses this region in time $\Delta t(q) \sim \xi/\vf \ll \ts = \ls/\vf$, so SRD and LRD act independently, and their scattering rates 
add according to Matthiessen's rule.


In the diffusive regime ($\xi \gg \ls$), scattering by SRD broadens electron states, requiring the energy-conserving delta function in FGR to be replaced by a Lorentzian of width $1/\ts$~\cite{Geldart:1975,Entin:1975,Geldart:1985,Kataoka:2001}. Consequently, $\Delta t(q) \sim \ts = \text{const}$, and the integrand in Eq.~\eqref{FGR} becomes \emph{less} singular as $q \to 0$, thus weakening the dGF anomaly. However, the dominant effect of diffusion is that it significantly 
enhances
the time required for an electron to traverse a region of size $\sim \xi$, now given by $\Delta t(q) \sim 1/D_{\rm s} q^2$, where $D_{\rm s} = \vf^2 \ts / d$ is the diffusion coefficient due to SRD. As a result, the integrand in Eq.~\eqref{FGR} becomes \emph{more} singular than in the ballistic regime: the $1/q^2$ divergence in $\Delta t(q)$ cancels the $q^2$ transport factor, reducing the integral to $\int dq\, q^{d-1} W(q) \sim \langle \delta M^2 \rangle$. Thus, $\delta\sigma_\ell \sim \delta\sigma_{\rm fl} \propto \langle \delta M^2 \rangle$, consistent with Eq.~\eqref{SM}.

\begin{figure}
[!]
\begin{centering}
\includegraphics[width=1.0
\columnwidth]
{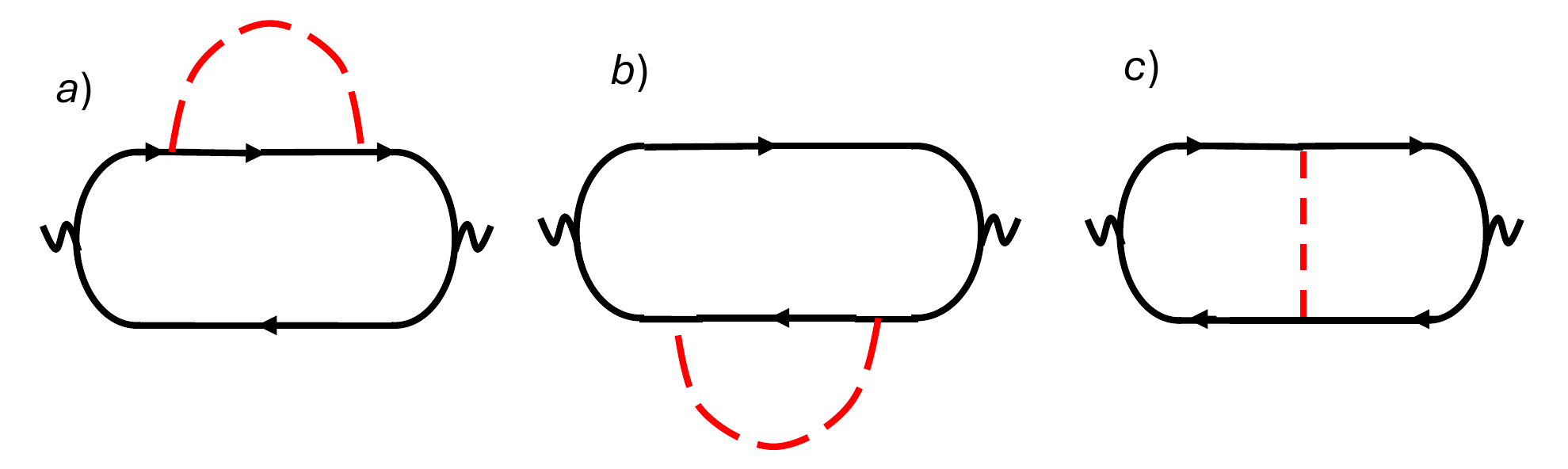}
\includegraphics[width=1.0
\columnwidth]
{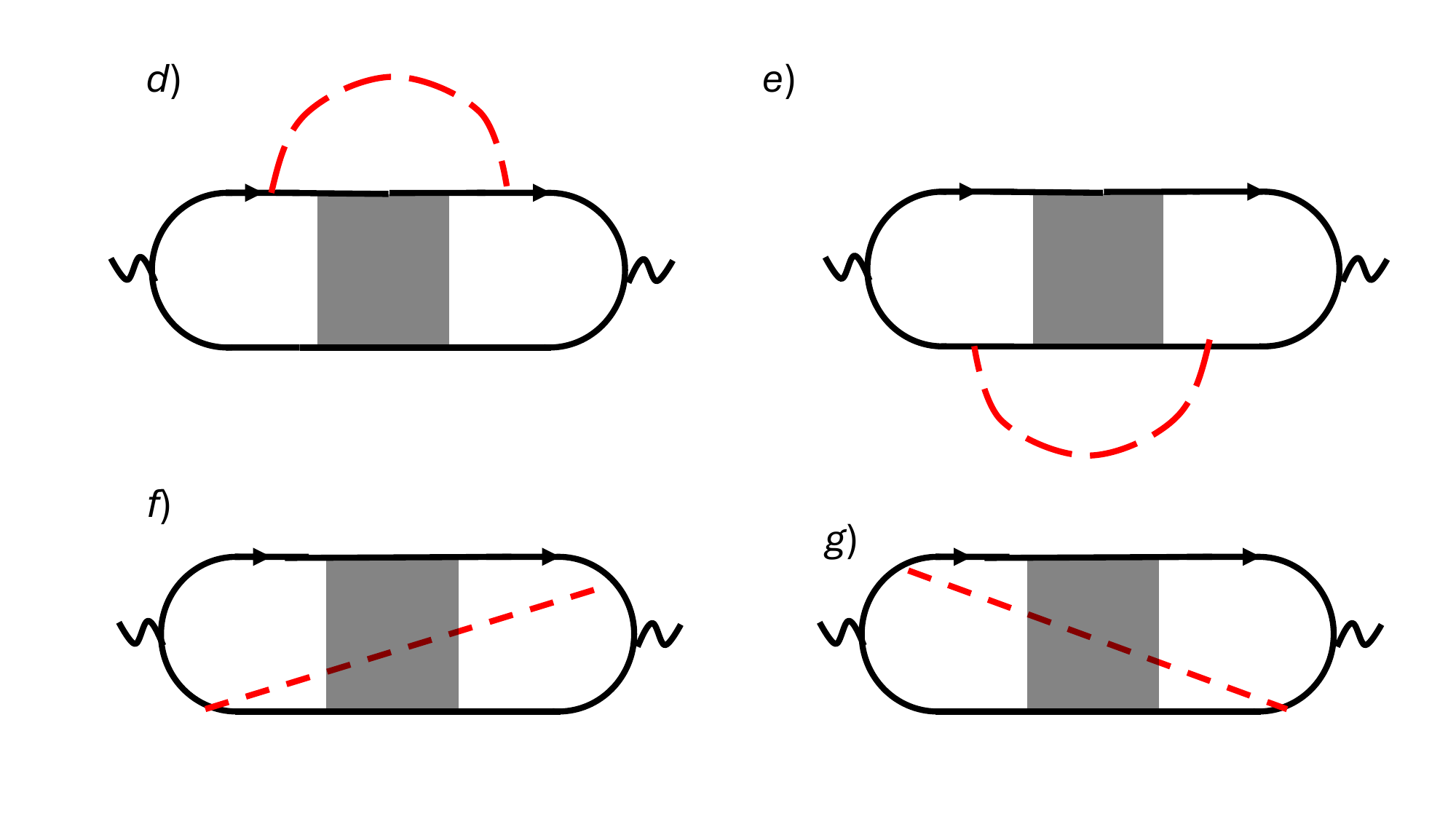}
\includegraphics[width=1.0
\columnwidth]
{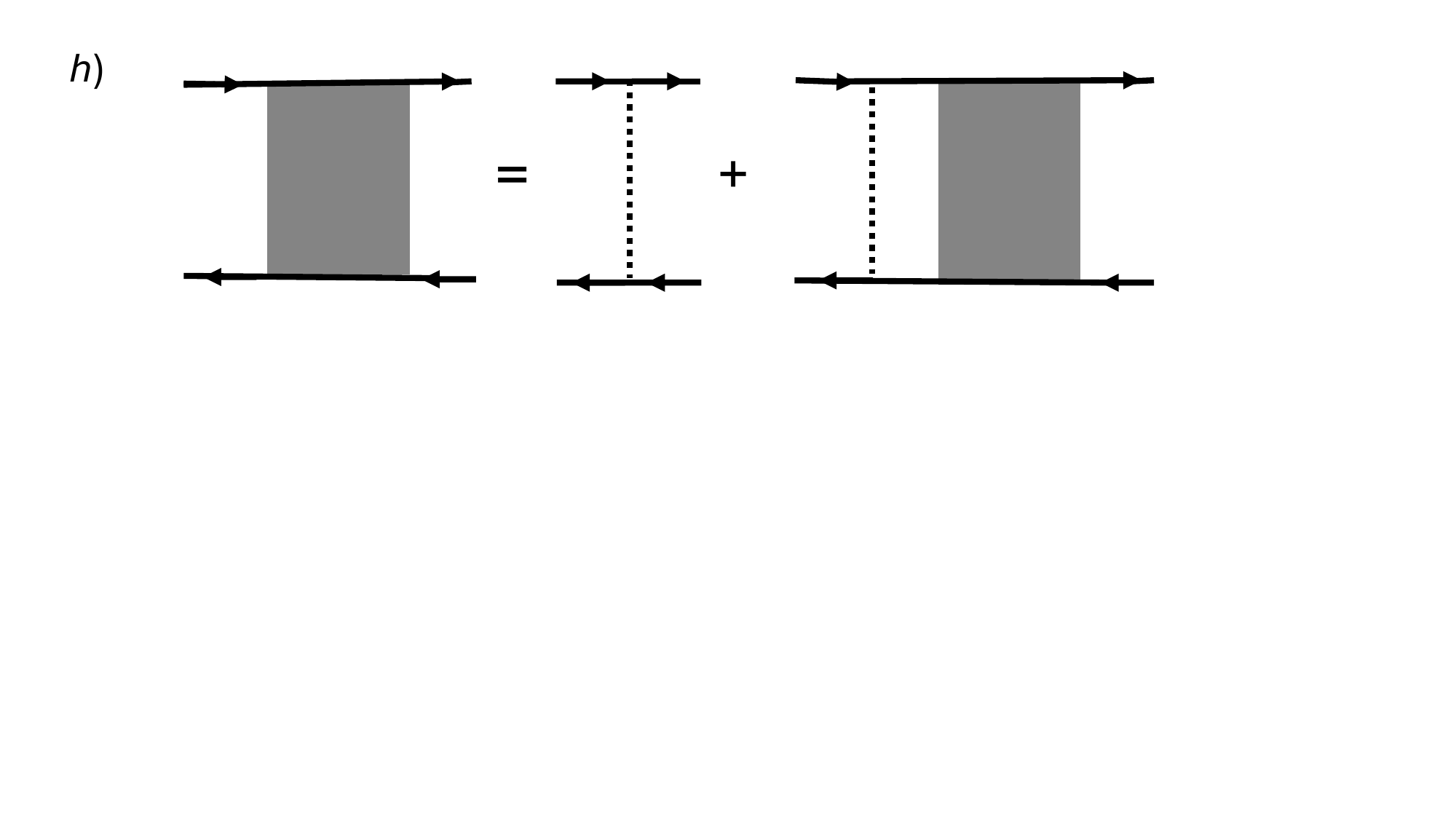}
\par\end{centering}
\vspace{-1in}
\caption{ 
Corrections to the  Kubo part of the measured conductivity, $\sigma_{\rm K}$ in Eq.~\eqref{mes}, due to long-range disorder. 
Solid lines:  Green's functions  
averaged over realizations of short-range disorder; wiggly lines: current vertices, dashed line: correlation function of long-range disorder; dotted line: correlation function of short-range disorder; shaded box: diffuson ladder, satisfying the equation shown graphically by diagram \emph{h}. 
\label{fig:SEMT}}
\end{figure}

\paragraph{Quantitative analysis.}  

We now outline the main steps of the derivation. We consider a 
metal with 
%
parabolic dispersion $\ve_\bk = k^2/2m - \mu$, subject to the condition $\mu\ts\gg 1$, where $\mu$ is the chemical potential. The leading corrections to the Kubo term in Eq.~\eqref{mes} due to 
LRD
are captured by diagrams \emph{a--g} in Fig.~\ref{fig:SEMT}. Solid lines represent disorder-averaged retarded/advanced Green’s functions,
\bea
G^{R/A}_\bk(\omega) = \frac{1}{\omega - \ve_\bk \pm i\ts/2},
\eea
while dashed and dotted lines correspond to the correlation functions of LRD and 
SRD,
respectively. The shaded box denotes the diffuson ladder $\mathcal{L}^R(q,\omega)$, which satisfies the integral equation shown in panel \emph{h}. Diagrams \emph{a--c} were previously analyzed in Ref.~\cite{Takada:1974}, but only for $Q\ls \gg 1$, which excludes the FM case. Diagrams \emph{d--g}
were shown in Ref.~\cite{Wilke:2000} to 
reflect a classical memory effect in the optical conductivity: a power-law, rather than exponential, decay of the velocity-velocity correlation function at long times~\cite{ERNST:1971,*Hauge:1974}.


In general, the correlator  of magnetic fluctuations in a FM 
can be written as
\bea
W(q) = q^{-2+\eta} F(q\xi), \label{Wqgen}
\eea
where $F(x)$ is a universal scaling function. Using the $\epsilon$-expansion and renormalization group, its asymptotic form for $q\xi \gg 1$ 
was found to be~\cite{Fisher:1973,Wegner:1975}:
\bea
F(q\xi \gg 1) = A + B_{\pm}\, \text{sgn}\,\theta \left( \frac{|\theta|}{q^{1/\nu}} \right)^{1-\alpha} + C_{\pm} \frac{|\theta|}{q^{1/\nu}} + \dots, \label{wegner}
\eea
with $A > 0$, and $B_{\pm}$, $C_{\pm}$ generally different
above and below $T_c$. 
The critical exponents $\alpha$, $\eta$, and $\nu$ for common universality classes are listed in Table~\ref{table:exponents}.

The sum of diagrams \emph{a--c} in Fig.~\ref{fig:SEMT}
gives the following  correction to the \emph{dc} conductivity 
\bea
\delta\sigma_{\ell,a\text{--}c} = -\frac{e^2}{2\pi d m^2} \int_\bq q^2 W(q) \int_\bk \left| G_{\bk - \bq/2}^R(0) G_{\bk + \bq/2}^R(0) \right|^2, \label{sigma_ac}
\eea
where $\int_\bp \equiv \int d^d p / (2\pi)^d$. The 
overall $q^2$ factor
 is the same transport factor as in Eq.~\eqref{FGR}. The momentum integral is evaluated using $\int_\bk = \nf \int d\ve_\bk \int_{\hat{k}}$, with $\nf$ the density of states at the Fermi level and $\int_{\hat{k}}$ denoting angular averaging. For $q \sim \xi^{-1} \ll \kf$, the dispersion simplifies to $\ve_{\bk \pm \bq/2} \approx \ve_\bk \pm \vf \hat{k} \cdot \bq / 2$, yielding
\bea
\delta\rho_{\ell,a\text{--}c}/\rs = -\delta\sigma_{\ell,a\text{--}c}/\sigma_{\rm s} = (\ts^2/\kf^2) \int_\bq q^2 W(q) f_d(q\ls), \label{ac}
\eea
with $\rs = 1/\sigs$, $f_2(x) = 1/\sqrt{x^2 + 1}$, and $f_3(x) = \tan^{-1}x/x$. 
For $q \gg \ls^{-1}$,
Eq.~\eqref{ac} is reduced back to FGR, Eq.~\eqref{FGR}. The FL result [Eq.~\eqref{FL}] arises from 
the second term in Eq.~\eqref{wegner}, 
and is present in both ballistic and diffusive regimes. Subtracting it off, the remaining contribution from $q \sim \xi^{-1}$ scales as $\delta\rho_{a\text{--}c} \propto |\theta|^{(d-1+\eta)\nu}$,
which 
generalizes
the dGF result beyond the mean-field level.
In this limit, Matthiessen's rule holds: $\delta\rho_{a\text{--}c} = \rs (\ts/\tau_{\rm tr})$. In the diffusive regime ($q \ll \ls^{-1}$), $f_d(x) \approx f_d(0) = 1$, which adds
an extra $q$ factor to the integrand and 
yields
$\delta\rho_{a\text{--}c} \propto |\theta|^{(d + \eta)\nu}$, thus confirming that the dGF anomaly is weakened compared to the ballistic case.

Diagrams \emph{d--g} in Fig.~\ref{fig:SEMT}, which are of primary interest here, yield
\bea
\delta\sigma_{\ell,d-g}=-\frac{4 e^2}{\pi d}\int_\bq W(q)\mathcal{L}^R(q,0) \left(\im\,{\bf u}_\bq\right)^2,\label{sigmadg}
\eea
where $
{\bf u}_\bq=\int_\bk (\bk/m)\left\vert G^R_\bk(0)\right\vert^2 G^A_{\bk+\bq}(0)
$
is the current vertex and $\mathcal{L}^R(q,0)$ is the static diffuson ladder, given by
$\mathcal{L}^R(q,0)=(1/2\pi\nf\ts)\mathcal{D}_d(q\ls)$ 
with 
$\mathcal{D}_{2}(x)=\left(1-1/\sqrt{x^2+1}\right)^{-1}$ and $\mathcal{D}_{3}(x)=\left(1-\tan^{-1}x/x\right)^{-1}$. 
The limit
 $\mathcal{L}^R(q\to 0,0)\propto 1/q^2$ 
corresponds to the diffusive limit of the interaction time, $\Delta t(q)$ in Eq.~\eqref{FGR}.
Integrating over $\bk$, we obtain 
\bea
\frac{\delta\rho_{\ell,d-g}}{\rs}=-\frac{\delta\sigma_{\ell,d-g}}{\sigma_{\rm s}}=\frac{1}{4\mu^2}\int_\bq W(q)\mathcal{D}_d(q\ls) g_d(q\ls).\label{dg}
\eea
where $g_2(x)=x^2/(1+x^2)^3$ and $g_3(x)=x^2/(1+x^2)^2$.
In the ballistic limit, $\delta\rho_{\ell,d-g}\ll \delta\rho_{\ell,a-c}$. 
In the diffusive limit, one  can replace $\mathcal{D}_d(x)$ and $g_d(x)$ in Eq.~\eqref{dg} by their small-$x$ limits, i.e., by $d/x^2$ and $x^2$, respectively, which yields
\bea
\delta\rho_{\ell,d-g}/\rs=(d/4\mu^2)\int_\bq W(q)=(d/4)\langle V^2\rangle/\mu^2,\label{dg-diff}
\eea
where $\langle V^2\rangle=\int_\bq W(q)$ is the variance
of the potential energy due to LRD. $\delta\rho_{\ell,d-g}$ is obviously more singular than $\delta\rho_{\ell,a-c}$ in the diffusive limit, and thus the latter can be neglected. 

Finally, we come to the second, fluctuational term in Eq.~\eqref{mes}, which becomes relevant in the diffusion limit, when a region of size $\sim \xi$ can be assigned its own conductivity. 
At fixed $\mu$, fluctuations of the local conductivity result from 
fluctuations of the Fermi energy. 
If $\sigs\left[\ef(\br)\right]=\sigs\left[\mu-V(\br)\right]$ is the local conductivity due to SRD at fixed Fermi energy, then $\langle\delta
\sigma_{\rm fl}^2\rangle=\langle\sigs^2\left[\mu-V(\br)\right]\rangle-\langle \sigs\left[\mu-V(\br)\right]\rangle^2\approx \langle V^2\rangle
\left(\partial\sigs(\mu)/\partial\mu\right)^2
=\left(\langle V^2\rangle/\mu^2\right)
\sigs^2$, where at the last step we took into account that $\nu(\e)\ts(\e)={\rm const}$ for SRD. Therefore, the fluctuational correction 
equals to $\delta\rho_{\rm fl}/\rho_{\rm s}=-\delta\sigma_{\rm fl}/\sigs=\langle V^2\rangle/d\,\mu^2
$, which is of the same order as the Kubo contribution 
in Eq.~\eqref{dg-diff}. 
Adding up the Kubo and fluctuational corrections gives the final result in the diffusive limit:
\bea
\delta\rho/\rho_{\rm s}=(d/4+1/d)\langle V^2\rangle/\mu^2.\label{deltarhofin}
\eea

Substituting Eq.~\eqref{wegner} into Eq.~\eqref{deltarhofin}, integrating term by term over the interval $1/\xi\lesssim q\lesssim 1/\ls$, and keeping only the contribution from the lower limit~\footnote{Here, we assume the most common scenario in which the integral of $W(q)$ up to the upper cutoff $1/l_s$ remains far from saturating the sum rule $\int_{\rm BZ} W(q)\, d^dq = N S^2$, where $N$ is the total number of spins.}, we arrive at the scaling form in Eq.~\eqref{SM}:
\bea
\delta\rho\propto A'|\theta|^{2 \beta}
=A'|\theta|^{1-\alpha-\zeta }, 
\label{deltarho2}
\eea
where $2 \beta= (d-2+\eta)\nu >0$, $\zeta=(2-\eta)\nu-1$, and where 
the hyperscaling relation $\nu d =2-\alpha$ \cite{Cardy} was employed at the last step. 
The dominant singular contribution in the diffusive regime scales like the Bragg peak intensity (square of the order parameter). As long as $\zeta>0$, the ``diffusive anomaly'' in Eq.~\eqref{deltarho2} is more singular than the FL one, Eq.~\eqref{FL}. Table \ref{table:exponents} confirms that this holds for the most common universality classes.
The sign of the diffusive anomaly--whether it 
manifests as
a cusp or anti-cusp in $\rho$--is non-universal, as it depends on the relative magnitudes of the constants $A$, $B_\pm$, and $C_\pm$ in Eq.~\eqref{wegner}.
In particular, $\delta\rho$ in Eq.~\eqref{deltarho2} 
exhibits a cusp at $\theta=0$ if $A'<0$ and an anti-cusp if $A'>0$.

\paragraph{Optical conductivity.} We now turn to the anomaly in the optical conductivity. Following Ref.~\cite{Wilke:2000}, we initially ignore
the effect of electron-electron interaction on diffusion, and also
focus on the regime of $\Omega\ts\ll 1$, when the frequency enters only through the diffuson as ${\mathcal L}^R(q,\Omega)=(1/2\pi\nu_F\ts^2)(\Ds q^2-i\Omega)^{-1}$,
while the rest of a diagram can be evaluated at zero frequency. In this regime, the dominant contribution to the optical conductivity is given by diagrams \emph{d-g}. At finite $\Omega$, Eq.~\eqref{sigmadg} is replaced by  
\bea
\re\delta\sigma(\Omega)=-\frac{4e^2}{\pi d} \int_\bq W(q) \re\mathcal{L}^R(q,\Omega)\left(\im\, {\bf u}_\bq\right)^2. 
\eea
In the diffusive limit, $(\im\, {\bf u}_\bq)^2=(\pi\nf\ts/\kf)^2(q\ls)^2$, 
which yields for
$\Delta\re\sigma(\Omega)\equiv\re\delta\sigma(\Omega)-\delta\sigma(0)$:
\bea
\Delta\re\sigma(\Omega)=\frac{d}{2}
\sgs
\frac{\Omega^2}{E_F^2}  \int_\bq \frac{ W(q)}
{(\Ds q^2)^2+\Omega^2},
\eea
where $\sgs=1/\rs$.
For $\Omega\ll \Ds/\xi^2$, one can set $q=0$ in $W(q)$ and factor it out from the integral, yielding
\bea
\Delta\re\sigma(\Omega)=
b_d
\sgs\left(\frac{|\Omega|}{\Ds}\right)^{d/2}\frac{W(0)}{\mu^2},
\eea
where $b_2=1/16$ and $b_3=\sqrt{2}/16\pi$.
For $\Ds/\xi^2\ll\Omega\ll 1/\ts$, one can neglect the 
$(\Ds q^2)^2 $   
term in the denominator, arriving at the $\Omega$-independent value:
$\Delta\re\sigma(\Omega)=-\sgs (d/4)\langle V^2\rangle/\mu^2$. Finally,  the conventional Drude behavior is recovered for $\Omega\gg 1/\tau_s$, i.e., $\re\sigma(\Omega)\propto 1/\Omega^2\ts$.
Note that a nonanalytic frequency dependence comes only from the Kubo part of the measured conductivity, whereas the fluctuational part gives the usual Drude behavior.

\begin{figure}[!]
\begin{centering}
\includegraphics[width=1.0
\columnwidth]{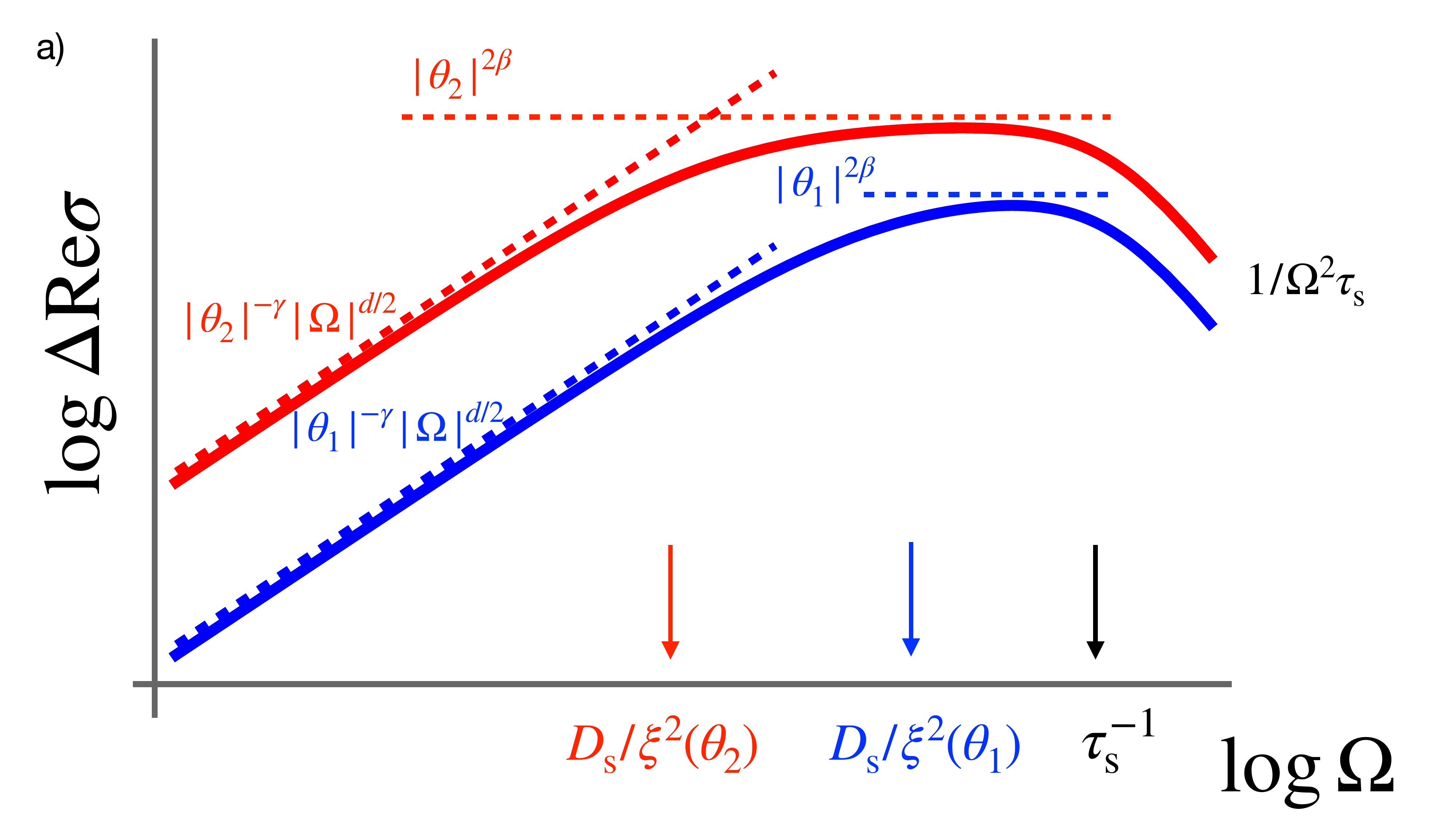}
\includegraphics[width=1.0
\columnwidth]{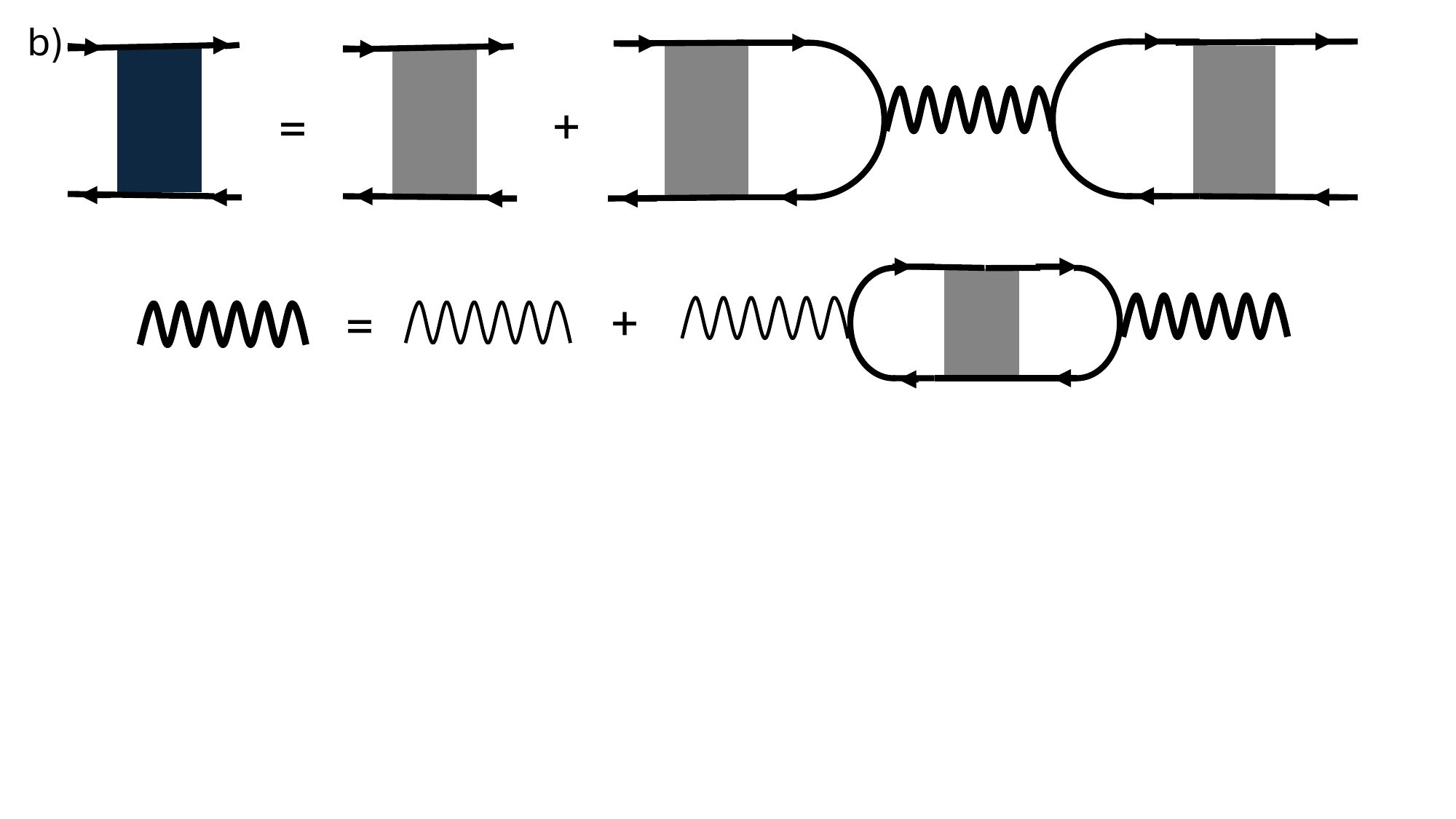}
\end{centering}
\vspace{-1.0in}
\caption{ 
(a) Non-Drude behavior of the optical conductivity due to a classical memory effect. 
Here, $\Delta\re\sigma = \re\delta\sigma(\Omega) - \delta\sigma(0)$, where $\delta\sigma(0)$ is the \emph{dc} correction 
due to LRD,
$\ts$ is the mean free time due to SRD, $\Ds$ the corresponding diffusion coefficient, and $\theta = (T - T_c)/T_c$. Two curves correspond to $|\theta_1| > |\theta_2|$. Slanted dashed lines show the non-analytic scaling $\Delta\re\sigma \propto |\Omega|^{d/2}$ for $\Omega \ll \Ds/\xi^2(|\theta|)$. Horizontal dashed lines mark the plateaus for $\Ds/\xi^2 \ll \Omega \ll 1/\ts$; the Drude behavior emerges for $\Omega \gg 1/\ts$. 
(b) Diagrammatic relation between screened (black box) and bare (shaded box) diffusons; thin and thick wavy lines denote bare and screened Coulomb interactions, respectively.
\label{fig:opt}}
\end{figure}
The behavior of the optical conductivity near a phase transition is sketched in Fig.~\ref{fig:opt} for two reduced temperatures, $|\theta|_1>|\theta|_2$. The slope of the nonanalytic part at lower frequencies  is proportional
to $W(0)\propto |\theta|^{-\gamma}$.
The non-analytic range is bounded from above by the Thouless energy at length $\xi$, $E_{\rm Th}(|\theta|)=\Ds/\xi^2(|\theta|)\propto |\theta|^{2\nu}$, which vanishes for $T \to T_c$. In the intermediate range, above $E_{\rm Th}(|\theta|)$ but below $1/\ts$, the optical conductivity reaches a plateau at a value that scales with $|\theta|$ is the same was as the {\emph dc} resistivity, i.e., as $|\theta|^{
2\beta}$. (We assume here 
that $\delta\rho$ in Eq.~\eqref{deltarho2} has a cusp at $\theta=0$).


As shown in Ref.~\cite{Evers:2001}, however, Coulomb interaction modifies electron diffusion. In the presence of dynamic screening, the bare diffuson (shaded box in Fig.~\ref{fig:opt}b) is replaced by the screened one, $\mathcal{L}^R_{\rm sc}(q,\omega)$ (black box), as depicted in the first line of panel b), where the thick wavy line represents the dynamically screened interaction $U^R_{\rm sc}(q,\omega)$. This is related to the bare interaction $U_0(q)$ via the random-phase approximation (second line). For $q\ls \ll 1$ and $\omega\ts \ll 1$, the screened diffuson reads
\bea
\mathcal{L}^R_{\rm sc}(q,\Omega)&=&\frac{1}{2\pi\nf\ts^2}\frac{1+\nf U_0(q)}{\Ds q^2\left(1+\nf U_0(q)\right)-i\Omega}.\label{scr_diff}
\eea
At $\Omega = 0$, this reduces to the singular static form $\propto 1/q^2$, implying no change to the \emph{dc} conductivity. For finite $\Omega$, and
with 
bare Coulomb interaction $U_0(q) \propto 1/q^{d-1}$, the denominator becomes $\Ds q^{3-d} \kappa_d^{d-1} - i\Omega$, which eliminates the diffusion pole and suppresses the non-analyticity in $\sigma(\Omega)$.\footnote{In $d=2$, a weak nonanalyticity survives: $\Delta\re\sigma(\Omega) \propto \Omega^2\ln\Omega^{-1}$ for $\Omega \ll \Ds\kappa_2/\xi$.} Here, $\kappa_d$ denotes the inverse screening length in $d$ dimensions.


Nevertheless, we propose two scenarios in which non-analytic behavior in $\sigma(\omega)$ may still be observable. The first, originally proposed in Ref.~\cite{Evers:2001}, involves a $d=2$ electron system screened by a metallic gate. The gate transforms the bare Coulomb potential into a dipolar one, yielding $U_0(q \to 0) = \text{const}$, so screening leads only to an irrelevant renormalization of the diffusion coefficient. We therefore suggest measuring $\sigma(\Omega)$ near a ferromagnetic transition in a gated 2D metal, such as Fe$_{3-x}$GeTe$_2$.

The second scenario involves a ferroelectric transition in a weakly doped insulator, where the divergence of the lattice dielectric constant, $\epsilon_{\rm L}$, near $T_c$ suppresses Coulomb interaction among itinerant electrons. For this mechanism to be effective, the inverse screening length must vanish faster than the correlation length. In $d=3$, $\kappa_3 \propto |\theta|^{\gamma/2}$ and $\xi \kappa_3 \propto |\theta|^{\gamma/2 - \nu}$, which only vanishes if $\gamma > 2\nu$---a condition marginally satisfied for 3D Ising transitions~\cite{Chaikin:book}. In $d=2$, however, $\kappa_2 \propto |\theta|^{\gamma}$, so $\xi \kappa_2 \to 0$ requires only $\gamma > \nu$, which holds for the 2D Ising class. We thus propose probing the optical conductivity near the phase transition in weakly doped 2D ferroelectrics such as $\alpha$-In$_2$Se$_3$, which undergoes a second-order transition in both bulk~\cite{Zheng:2018} and exfoliated forms~\cite{Zhou:2017}, and can be doped~\cite{Zaslonkin:2007}, or in monolayer SnTe~\cite{Kai:2016}. At low doping, carriers near $T_c$ are likely to be non-degenerate, but this does not affect the resistive anomaly, which remains a single-particle effect. Our results extend to this regime by replacing $\ef \to T$.

\begin{table}[!]
\begin{ruledtabular}
 \begin{tabular}{c|c|c|c|c|c|c}
Universality class   & $\nu$ & $\eta$ &  $\alpha$  & $2\beta$ & $\gamma$ & $\zeta$\\
 \hline
 O(3), $d=3$~\cite{Campostrini02} & 0.71& 0.038 & -0.13& 0.738  & 1.40 & 0.39\\
 \hline
O(2), $d=3$~\cite{Campostrini01} & 0.67 & 0.038 & -0.015&  0.698 &1.32 &0.31\\ 
 \hline
 Ising, $d=3$~\cite{Kos16,Reehorst22} & 0.63 & 0.037  & 0.11& 0.652   &  1.24 &0.24 \\
 \hline
Ising, $d=2$ \cite{Cardy}& 1 & 1/4  & 0&  1/4  &  7/4 &3/4\\
\end{tabular}
 \end{ruledtabular}
\caption{
Critical exponents for common universality classes. $\nu$ governs the correlation length, $\eta$ is defined in Eq.~\eqref{Wqgen}, $\alpha$ is the specific heat exponent, $\beta$ describes the order parameter ($2\beta$ governs the Bragg peak intensity), $\gamma$ is the susceptibility exponent, and $\zeta = \nu(2 - \eta) - 1$. For the $d=2$ Ising model, $\alpha = 0$ implies a logarithmic divergence. Exponents for $d=3$ are rounded to two significant digits.
\label{table:exponents}}
\end{table}


\paragraph{Connection to the experiment.} 

While resistive anomalies near  $T_c$ are widely observed in metallic ferromagnets~\cite{Gerlach1932,BittelGerlach1938,Craig:1967,Campbell:1982}, the scaling predicted by Fisher and Langer (FL) is not universally obeyed. In some cases, such as Ni, the anomaly aligns with FL scaling over a finite temperature range~\cite{Kallback81}, whereas in others, like the rare-earth FMs RNi$_5$ (R = Tb, Dy, Er), $\rho$ itself—not $d\rho/dT$—peaks at $T_c$, deviating from FL behavior~\cite{Blanco1994}. Even for Ni, fitting the data using only FL singular and regular terms fails within $\pm1$ K of $T_c$~\cite{Kallback81}.

In contrast to the FL anomaly, which produces a peak in $d\rho/dT$, the diffusive mechanism discussed in this Letter yields a cusp or anticusp in $\rho$, consistent with the resistivity peak observed in RNi$_5$~\cite{Blanco1994}. A similar cusp in $\rho$, together with an asymmetric peak-antipeak structure in $d\rho/dT$ from the sum of regular and singular contributions, may also explain the anomaly seen in GaR$_2$ FMs (R = Ni, Rh, Pt)~\cite{Mydosh:1970}.

Resolving the FL and diffusive contributions may require measurements too close to $T_c$ for conclusive interpretation. A more robust signature is the non-Drude behavior of the optical conductivity, shown in Fig.~\ref{fig:opt}, which reflects classical memory effects. However, such behavior is only expected in gated $d=2$ systems or near ferroelectric
transitions.

An alternative route is to study periodic intermetallic magnets, where the spacing between localized moments matches $\ls$, thus extending the temperature range of the diffusive regime near $T_c$. These systems may permit extraction of the $\beta$ exponent from transport and its comparison to neutron scattering data. Ongoing efforts also aim to realize 2D analogs of such materials~\cite{Zhang2010}.

{\it Note added in proof:} Recent non-linear optical experiments on ferromagnetic
Ca$_2$RuO$_4$ (Bhandia et al., arXiv:2412.08749) confirmed that scattering by magnetic moments is elastic
and revealed a cusp in the momentum relaxation rate, consistent with our prediction.

We thank P. Armitage, G. Blumberg, A. Chubukov, V. Kravtsov, S. Kundu, A. Levchenko, S. Kivelson, A. McLeod, A. Mirlin, B. Shklovskii, Y. Wang, and K. Alp Yay for illuminating discussions.  We are especially grateful to D. Polyakov for bringing Ref.~\cite{Evers:2001} to our attention. DLM  acknowledges support from the National Science Foundation (NSF)  via grant  DMR-2224000, from the
Simons Foundation Targeted Grant 920184 to the W. I. Fine Theoretical Physics Institute, University of Minnesota, and
 hospitality of the  Kavli Institute for Theoretical Physics, Santa Barbara, supported by the NSF grants PHY-1748958 and PHY-2309135, 
 VIY acknowledges support from the Basic 
 Research
 Program of HSE.
 DLM and CDB thank the Aspen Center for Physics, supported by NSF grant PHY-2210452.
 CDB also acknowledges support from the Center for Nonlinear Studies (CNLS) through the Stanislaw M. Ulam Distinguished Scholar position, funded by LANL's LDRD program.


\bibliography{dm_references}
\newpage
\onecolumngrid
\setcounter{equation}{0}
\begin{center}
{\bf End Matter}
\end{center}
\section{Details of diagrammatic calculations}
\subsection{Diagrams \emph{a-g}, Fig.~\ref{fig:SEMT}}
The Kubo formula for the conductivity at frequency $\Omega$ reads
\bea
\re\sigma(\Omega)=\frac{\im K^R(\Omega)}{\Omega},
\eea
where the retarded current-current correlation function is obtained from its Matsubara counterpart via analytic continuation 
\bea
 K^R(\Omega)=K(i\Omega_n\to i\Omega+0^+)=\frac{1}{d}\int^{1/T}_0 d\tau e^{i\Omega_n\tau}
 \langle {\bf j}(\tau){\bf j}(0)\rangle\Big\vert_{i\Omega_n\to i\Omega+0^+}.
\eea
\paragraph{Diagrams a-c, Fig.~\ref{fig:SEMT}.}
 Diagrams \emph{a-c} for $K(i\Omega_n)$ read
\bse
\bea
K_{ab}(i\Omega_n)&=& K_a(i\Omega_n)+K_b(i\Omega_n)=-\frac{2e^2}{d m^2}T\sum_{\omega_m} \int_\bk k^2 \left[G^2_\bk(i\omega_m+i\Omega_n)\Sigma_{\bk}(i\omega_m+i\Omega_n) G_\bk(i\omega_m)\right.\nn\\
&&\left.+G^2_\bk(i\omega_m)\Sigma_\bk(i\omega_m)G_{\bk}(i\omega_m+i\Omega_n)\right],\label{Kab}\\
K_{c}(i\Omega_n)&=& K_c^{(1)}+K_c^{(1)},\label{Kc}\\
K_c^{(1)}&=&-\frac{2e^2}{d m^2}T\sum_{\omega_m}\int_{\bk,\bq} k^2 G_\bk(i\omega_m+i\Omega_n)G_{\bk+\bq}(i\omega_m+i\Omega_n) G_{\bk+\bq}(i\omega_m)G_\bk(i\omega_m)W(q),\label{Kc1}\\
K_c^{(2)}&=&-\frac{2e^2}{d m^2}T\sum_{\omega_m}\int_{\bk,\bq} (\bk\cdot\bq) G_\bk(i\omega_m+i\Omega_n)G_{\bk+\bq}(i\omega_m+i\Omega_n) G_{\bk+\bq}(i\omega_m)G_\bk(i\omega_m)W(q),\label{Kc2}
\eea
\end{subequations}
where $
G_\bk(i\omega_m)=\left(i\omega_m-\varepsilon_\bk+i\rm{sgn}\omega_m/2\ts\right)^{-1}$
and 
$
\Sigma_\bk(i\omega_m)=\int_{\bq} G_{\bk+\bq}(i\omega_m)W(q)$
is the lowest-order self-energy due to scattering by LRD.
Without loss of generality, we choose $\Omega_n>0$ such that the integral over $\varepsilon_\bk$ is non-zero only if $\omega_m<0$ and $\omega_m+\Omega_n>0$.  
Applying several times the identity
$G_\bk(i\omega_m) G_\bk(i\omega_m + i \Omega_n) =
\left[G_\bk(i\omega_m) -G_\bk(i\omega_m + i \Omega_n)\right]
/i(\Omega_n+1/\ts)
$ to Eqs.~\eqref{Kab} and \eqref{Kc1}, we find that they cancel each other:
\bea
K_{ab}(i\Omega_n)=-K^{(1)}_{c}(i\Omega_n)
&=&
\frac{2e^2}{d m^2}\frac{1}{(\Omega_n+1/\ts)^2}T\sum_{\omega_m}\int_{\bk,\bq} k^2\left[G_\bk(i\omega_m)G_{\bk+\bq}(i\omega_m+i\Omega_n)+G_\bk(i\omega_m+i\Omega_n)G_{\bk+\bq}(i\omega_m)\right]W(q).\nn\\
\eea

Although the remaining part, $K_{c2}$, appears to be linear in $q$, one can show, by relabeling the momenta as 
 $\bk\to\bk-\bq/2$ and $\bk+\bq\to \bk+\bq/2$,
that it is, in fact, quadratic in $q$:
\bea
K_{c2}=-\frac{2e^2}{d m^2}\frac{1}{(\Omega_n+1/\ts)^2}T\sum_{\omega_m}\int_{\bk,\bq}&& q^2 G_{\bk-\bq/2}(i\omega_m+i\Omega_n)G_{\bk+\bq/2}(i\omega_m+i\Omega_n) G_{\bk+\bq/2}(i\omega_m)G_{\bk-\bq/2}(i\omega_m)W(q).
\eea

Carrying out analytic continuation, discarding $G^R G^R$ and $G^A G^A$ parts, and taking the limit $\Omega\to 0$, we arrive at Eq.~\eqref{sigma_ac} of the Main Text (MT).

Next, we replace $d^dk/(2\pi)^d$ by $\nf\!\!\int d\varepsilon_\bk\int d\mathcal{O}_\bk/\mathcal{O}_d$, where $\nf$ is the density of states per spin at the Fermi energy, $d\mathcal{O}_\bk$ is the solid angle subtended by $\bk$, $\mathcal{O}_2=2\pi$, and $\mathcal{O}_3=4\pi$, and expand $\e_{\bk\pm\bq/2}=\varepsilon_\bk\pm\vf \hat k\cdot\bq/2$, where $\hat k=\bk/k$. After integration over $\ve_\bk$ and $\mathcal{O}_\bk$, we obtain
\bea
\int_\bk \left\vert G^R_{\bk-\bq/2}(0)G^R_{\bk+\bq/2}(0)\right\vert^2
=-4\pi \nf\ts^3\, \im\int\frac{d\mathcal{O_\bk}}{\mathcal{O}_d}\frac{1}{i-\hat k\cdot\bq\ls}=4\pi \nf\ts^3
\left\{
\begin{array}{ccc}
(q^2\ls^2+1)^{-1/2},\;d=2\\
\tan^{-1}(q\ls)/q\ls,\;d=3\\
\end{array}
\right.\equiv 4\pi \nf\ts^3f_d(q\ls).
\eea

\paragraph{Diagrams d-g, Fig.~\ref{fig:SEMT}.}
Diagrams \emph{d}-\emph{g}  can be calculated immediately in the static limit, when \emph{d} and is equal to \emph{e}, and  \emph{f} is equal to\emph{g}.
Carrying out analytic continuation, taking the static limit, and discarding the $G^RG^R$ and $G^AG^A$ parts, we obtain Eq.~\eqref{sigmadg} of MT.

In an isotropic system, ${\bf u}_\bq$ must be collinear with $\bq$. Dotting  ${\bf u}_\bq$ into $\hat q=\bq/q$, we obtain with $\e\equiv\varepsilon_\bk$
\bea
\im\left( {\bf u}_{\bq}\cdot\hat q\right)=
\frac{1}{m}
\int\frac{d\mathcal{O}_\bk}{\mathcal{O}_d}\int d\e\nu(\e) k(\e)
\hat k\cdot \hat q\left\vert G^R_\bk(0)\right\vert^2 
\frac{1/2\ts}{\left[\e+\varv(\e)q
\hat k\cdot \hat q\right]^2+(1/2\ts)^2}.\label{imu}
\eea
Unlike the case for diagrams \emph{a-c}, 
projecting the integrand of the last equation onto the Fermi surface, i.e, putting $\nu(\ve)=\nu_F$, $k(\ve)=\kf$, and $\varv(\ve)=\vf$,  gives a zero result \cite{Wilke:2000}, because
the integrand becomes odd under $\e\to-\e$ and $\hat k\to-\hat k$ on such projection.
To obtain a non-zero result, one needs to expand $\nu(\ve)=\nf+\nf'\e$, $k(\ve)=\kf+\e/
\vf$, and $\varv(\ve)=\vf+\ve/\kf$, where $\nf'=\partial\nu/\partial \ve\vert_{\ve=\ef}=(d/2-1)\nf/\ef$ for a parabolic spectrum in $d$ dimensions. Integrating over $\e$ and $\mathcal{O}_\bk$, we obtain
\bea
\im\left( {\bf u}_{\bq}\cdot \hat q\right)
&=&
\frac{\pi \nf\ts}{\kf}\re\left[(d-1) 
\int\frac{d\mathcal{O}_\bk}{\mathcal{O}_d} \frac{\cos\theta}{i+q\ls\cos\theta}
-q\ls
\int\frac{d\mathcal{O}_\bk}{\mathcal{O}_d} \frac{\cos^2\theta}{\left(i+q\ls\cos\theta\right)^2}
\right],\nn\\
&=&
\pi \frac{q\nf\ts^2}{m}
\times\left\{
\begin{array}{cc}
&1/\left[(q\ls)^2+1\right]^{3/2},\;d=2;\\
&\\
&1/\left[(q\ls)^2+1\right],\;d=3.   
\end{array}
\right.
\label{Vpp}
\eea
Upon taking a square, the last result reproduces the expression  for $(\im\,\bu_\bq)^2$ quoted in the MT.
\subsubsection{Diagrams in panel b), Fig.~\ref{fig:opt}}
In the diffusive limit, the dynamically screened potential of Coulomb interaction is given by \cite{altshuler:1985}
\bea
U^R_{\rm sc}(q,\Omega)=\frac{U_0(q)}{1+U_0(q)\nf\frac{\Ds q^2}{\Ds q^2-i\Omega}}.\label{Udyn}
\eea
Algebraically, the diagram for the screened diffuson reads
\cite{Evers:2001}
\bea
\mathcal{L}^R_{\rm sc}(q,\omega)&=&\mathcal{L}^R(q,\omega)\left[1-2\pi i \Omega (\nf\ts)^2\mathcal{L}^R(q,\Omega) U^R_{\rm sc}(q,\Omega)
\right].
\label{Ldyn}
\eea
Substituting  Eq.~\eqref{Udyn} into  Eq.~\eqref{Ldyn}  yields Eq.~\eqref{scr_diff} of the MT.
\section{Integral of the spin-spin correlation function beyond the mean-field level}
In this section we show how Eq.~\eqref{deltarho2} for the resistive anomaly in the diffusive regime was obtained.
Substituting Eq.~\eqref{Wqgen} with $F(q\xi)$ given by Eq~\eqref{wegner} into Eq.~\eqref{dg-diff}, and limiting the range of integration to $\xi^{-1}\propto |\theta|^\nu\lesssim q\lesssim 1/\ls$, we obtain
\bea
\delta\rho&\propto &\int^{1/\ls}_{ |\theta|^\nu}dq\left(Aq^{d-3+\eta}+B_{\pm} {\rm sgn}\theta\,|\theta|^{1-\alpha} q^{d-3+\eta-(1-\alpha)/\nu}+C_{\pm}|\theta|q^{d-3+\eta-1/\nu}\right).
\eea
In the second term, we neglect its contribution from the upper limit, because it is of same form ($\propto {\rm sgn\theta}\,|\theta|^{1-\alpha}$) but of smaller magnitude as the FL term, coming from the range of $q\sim \kf$ in the ballistic limit. Then,
\bea
\delta \rho&\propto &\left[ -\frac{A}{d-2+\eta}|\theta|^{(d-2+\eta)\nu}-{\rm sgn}\theta\,|\theta|^{1-\alpha}\frac{B_{\pm}}{d-2+\eta-(1-\alpha)/\nu}
|\theta|^{(d-2+\eta)\nu-(1-\alpha)}
-\frac{C_{\pm}}{d-2+\eta-1/\nu}|\theta|^{(d-2+\eta)\nu}\right]\nn\\
&=&A'|\theta|^{(d-2+\eta)\nu}=A'|\theta|^{2\beta},
\eea
where $\beta=(d-2+\eta)\nu/2$ is the critical exponent of the order parameter, defined by $\langle M\rangle\propto (-\theta)^\beta$ for $\theta<0$, and
\bea
A'&=&-\left[\frac{A}{d-2+\eta}+{\rm sgn}\theta \frac{B_{\pm}}{2-\eta-1/\nu}+\frac{C_\pm}{d-2+\eta-1/\nu}\right].
\eea
Using the hyperscaling relation $\nu d =2-\alpha$, the results can be re-written as
\bea
\delta\rho&=&A'|\theta|^{1-\alpha-\zeta}
\nn\\
A'&=&-\frac{A}{d-2+\eta}+{\rm sgn}\theta\frac{B_\pm}{2-\eta-1/\nu}-\frac{C_\pm}{d-2+\eta-1/\nu},\nn\\
\eea
where $\zeta=\nu(2-\eta)-1$.

\section{Alternative interpretation of the resistive anomaly}
As it was argued in the main text, the most adequate interpretation of the enhanced resistivity anomaly in the diffusive regime is a classical memory effect, 
arising
from multiple returns of the electron trajectory to the same position \cite{Wilke:2000}. However, we would also like to point out a similarity between diagrams \emph{d-g} in Fig.~\ref{fig:SEMT} and the diagrams describing the Altshuler-Aronov (AA) correction to the conductivity, which arises from quantum interference between electron-electron and electron-impurity interactions \cite{altshuler:1985}. In AA diagrams, the dashed line is replaced with the dynamical potential of electron-electron interaction and, in addition, each vertex of this interaction is dressed by a diffuson ladder. Since our LR potential is static, it cannot change the analyticity of the Green's functions adjacent to the vertex, and vertex corrections vanish to leading order in $1/\mu\ts$. However, the common theme of quantum interference and classical memory effects is that diffusing electrons are more susceptible to external perturbations, be it the Coulomb potential of other electrons in the AA case or LRD in our case.

\end{document}